\documentclass[10pt,final,journal,letter,oneside,twocolumn]{IEEEtran}

\usepackage[noadjust]{cite}
\usepackage{srcltx}
\usepackage{graphicx}
\usepackage{epstopdf}
\usepackage{psfrag}
\usepackage{epsfig}
\usepackage[tbtags,sumlimits,nointlimits,reqno]{amsmath}
\usepackage{amssymb}
\usepackage{xcolor}
\usepackage{float}
\usepackage{subfigure}
\usepackage{soul}
\usepackage{footnote}
\usepackage{amsmath}
\graphicspath{{Figs/}}

\begin{document}
\title{Enhanced Bandwidth and Diversity in Real-Time Analog Signal Processing (R-ASP) using Nonuniform C-section Phasers}
\author{Sajjad~Taravati,~\IEEEmembership{Student Member,~IEEE,}
                 Shulabh Gupta,~\IEEEmembership{Member,~IEEE,}
                 Qingfeng Zhang,~\IEEEmembership{Member,~IEEE,}
              and~Christophe~Caloz,~\IEEEmembership{Fellow,~IEEE}
\thanks{Sajjad Taravati, Shulabh Gupta and Christophe Caloz are with the Department of Electrical Engineering, Polytechnique Montr\'{e}al, Montr\'{e}al, QC, Canada H3T 1J4 (e-mail: sajjad.taravati@polymtl.ca).}
\thanks{Qingfeng Zhang is with the Electrical Engineering Department of South University of Science and Technology of China, Shenzhen, China.}
}

\markboth{IEEE Microwave and Wireless Component Letters,~Vol.~**, No.~**, **~2016}%
{Shell \MakeLowercase{\textit{et al.}}: Bare Demo of IEEEtran.cls for Journals}

\maketitle

\begin{abstract}
We show that a continuously nonuniform coupled-line C-section phaser, as the limiting case of the step-discontinuous coupled-line multisection commensurate and non-commensurate phasers, provides enhanced bandwidth and diversity in real-time analog signal processing (R-ASP). The phenomenology of the component is explained in comparison with the step-discontinuous using multiple-reflection theory and a simple synthesis procedure is provided. The bandwidth enhancement results from the suppression of spurious group delay harmonics or quasi-harmonics, while the diversity enhancement results from the greater level of freedom provided by the continuous nature of the nonuniform profile of the phaser. These statements are supported by theoretical and experimental results.
\end{abstract}


\IEEEpeerreviewmaketitle

\section{Introduction}
Real-time Analog Signal Processing (R-ASP) is a potential alternative to dominantly digital radio technology, given its high-speed, low-consumption and frequency-scalability benefits~\cite{Caloz_MM_RASP_Caloz_09_2013}. The key component of a R-ASP system is the phaser~\cite{Caloz_MM_RASP_Caloz_09_2013,Gupta_TMTT_03_2015}\footnote{The term ``phaser'' was explained in~\cite{Caloz_MM_RASP_Caloz_09_2013} and~\cite{Gupta_TMTT_03_2015}.}, a device providing specified group delay versus frequency response depending on the application, e.g. linear for real-time Fourier transformers~\cite{Laso_MTT_60_2012}, staircase for spectrum sniffers~\cite{Nikfal_MWCL_11_2012} and Chebyshev for dispersion code multiple access (DCMA)~\cite{Nikfal_Patent_2013,Taravati_EuMC_2014}.

Step-discontinuity coupled-line all-pass phasers represent a common type of phasers~\cite{Caloz_MM_RASP_Caloz_09_2013,Gupta_TMTT_03_2015,Schiffman_MTT_1955,Cristal_TMTT_1966}, but they suffer from bandwidth restriction, due to the presence of spurious group delay harmonics, and restricted dispersion diversity, in R-ASP. This paper shows how these limitations can be mitigated by using \emph{continuously nonuniform} C-section phasers~\cite{Taravati_EuMC_2014} as the subwavelength section limit of step-discontinuity multi-section coupled-line phasers~\cite{Cristal_TMTT_1966}.
\section{Step-Discontinuity Nonuniform Phasers}\label{sec:cascaded}
\begin{figure}
\centering
{\psfrag{a}[c][c][0.95]{$d_1$}
\psfrag{b}[c][c][0.95]{$d_2$}
\psfrag{c}[c][c][0.95]{$d_3$}
\psfrag{A}[c][c][0.95]{$1$}
\psfrag{B}[c][c][0.95]{$2$}
\psfrag{C}[c][c][0.95]{$3$}
\psfrag{d}[c][c][0.95]{$d_{M}$}
\psfrag{e}[c][c][0.95]{$D$}
\psfrag{x}[c][c][0.95]{$v_\text{in}$}
\psfrag{g}[c][c][0.95]{$D$}
\psfrag{y}[c][c][0.95]{$v_\text{out}$}
\psfrag{w}[c][c][0.95]{$\Gamma_p^\text{in}$}
\psfrag{f}[c][c][0.95]{$\Gamma_{p,0}$}
\psfrag{g}[c][c][0.95]{$\Gamma_{p,1}$}
\psfrag{h}[c][c][0.95]{$\Gamma_{p,2}$}
\psfrag{i}[c][c][0.95]{$\Gamma_{p,3}$}
\psfrag{k}[c][c][0.95]{$Z^{\text{L}}_{p}$}
\psfrag{M}[c][c][0.95]{${M}$}
\centering{ \includegraphics[width=\columnwidth] {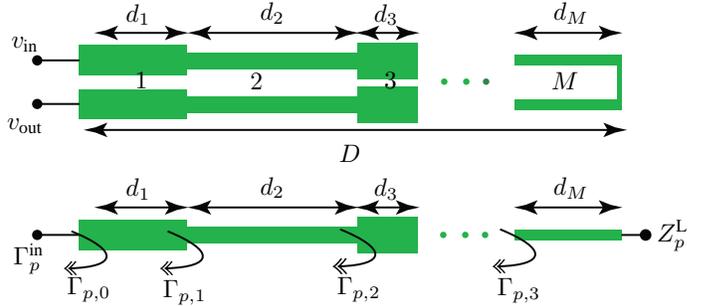}}}
\caption{Non-commensurate C-section phaser (top), and corresponding even- and odd-mode equivalent circuits (bottom).}
\label{Fig:gen_noncomm}
\end{figure}

\noindent Figure~\ref{Fig:gen_noncomm} shows a general step-discontinuity nonuniform coupled-line phaser with $M$ sub-sections of lengths $d_1, d_2, ..., d_M$ and corresponding even- and odd-mode equivalent circuits, denoted by the subscript $p$ ($p=$e,o resp.), where $Z^{\text{L}}_{\text{e}}=\infty$, $Z^{\text{L}}_{\text{o}}=0$.
For small discontinuities, the total even/odd reflection coefficients at the input are~\cite{Pozar_4th_2012}
\begin{equation}
\Gamma_p^\text{in}=\Gamma_{p,0}+\Gamma_{p,1} e^{-j2 \beta d_1}+...+\Gamma_{p,M} e^{-j2M \beta d_M},
\label{eqa:gama_in}
\end{equation}
where $\Gamma_{p,m}$ is the reflection coefficient between sections $m$ and $m+1$ and $\beta$ is the ($m$-independent, assuming TEM sections) guided wavenumber. The total transmission scattering parameter and group delay of the phaser follow as
\begin{equation}
S_{21}=\frac{1}{2}(\Gamma_\text{e}^\text{in}-\Gamma_\text{o}^\text{in})=\frac{1}{2}\sum_{m=0}^{M-1} \left(\Gamma_{\text{e},m}-\Gamma_{\text{o},m} \right) e^{-j2m \beta d_m},
\label{eqa:s21}
\end{equation}
\begin{equation}
\tau (\omega)=-\frac{d\phi_{S_{21}}}{d \omega}=\sum_{m=0}^{M-1} \left(\frac{-d \phi_{( \Gamma_{\text{e},m}- \Gamma_{\text{o},m})}}{d \omega}+  \frac{2m d_m}{v} \right),
\label{eqa:gd}
\end{equation}
where $v=\omega/\beta$ is the phase velocity.

\begin{figure}
\centering
{\psfrag{E}[c][c][0.8]{$d_1$}
\psfrag{F}[c][c][0.8]{$d_2$}
\psfrag{J}[c][c][0.8]{$d$}
\psfrag{H}[c][c][0.8]{$1$}
\psfrag{I}[c][c][0.8]{$2$}
\psfrag{G}[c][c][0.8]{$D$}
\psfrag{x}[c][c][0.8]{$v_\text{in}$}
\psfrag{y}[c][c][0.8]{$v_\text{out}$}
\psfrag{a}[c][c][0.8]{(a)}
\psfrag{b}[c][c][0.8]{(b)}
\psfrag{c}[c][c][0.8]{(c)}
\psfrag{A}[c][c][0.8]{$\beta D$}
\psfrag{B}[c][c][0.8]{Group delay (ns)}
\psfrag{C}[l][c][0.7]{$C_1=0.5, C_2=0.8$}
\psfrag{D}[l][c][0.7]{$C_1=0.1, C_2=0.9$}
\psfrag{e}[c][c][0.7]{$\pi$}
\psfrag{f}[c][c][0.7]{$2\pi$}
\psfrag{g}[c][c][0.7]{$3\pi$}
\psfrag{h}[c][c][0.7]{$4\pi$}
\psfrag{i}[c][c][0.7]{$5\pi$}
\psfrag{j}[c][c][0.7]{$6\pi$}
\psfrag{k}[l][c][0.7]{$C=0$}
\psfrag{l}[l][c][0.7]{$C=0.2$}
\psfrag{m}[l][c][0.7]{$C=0.4$}
\psfrag{n}[l][c][0.7]{$C=0.6$}
\psfrag{o}[l][c][0.7]{$C=0.8$}
\psfrag{M}[c][c][0.7]{$M=1$}
\psfrag{N}[c][c][0.7]{$M=2$}
\psfrag{q}[c][c][0.75]{\shortstack{$M=2$\\$d_1=0.58d_2$}}
\psfrag{R}[c][c][0.7]{$C$}
\psfrag{S}[c][c][0.7]{$C_1$}
\psfrag{s}[c][c][0.7]{$C_2$}
\psfrag{K}[c][c][0.7]{$\pi M$}
\psfrag{t}[c][c][0.7]{$BW_\text{max}(2\pi D/v)$}
\centering{ \includegraphics[width=0.8\columnwidth] {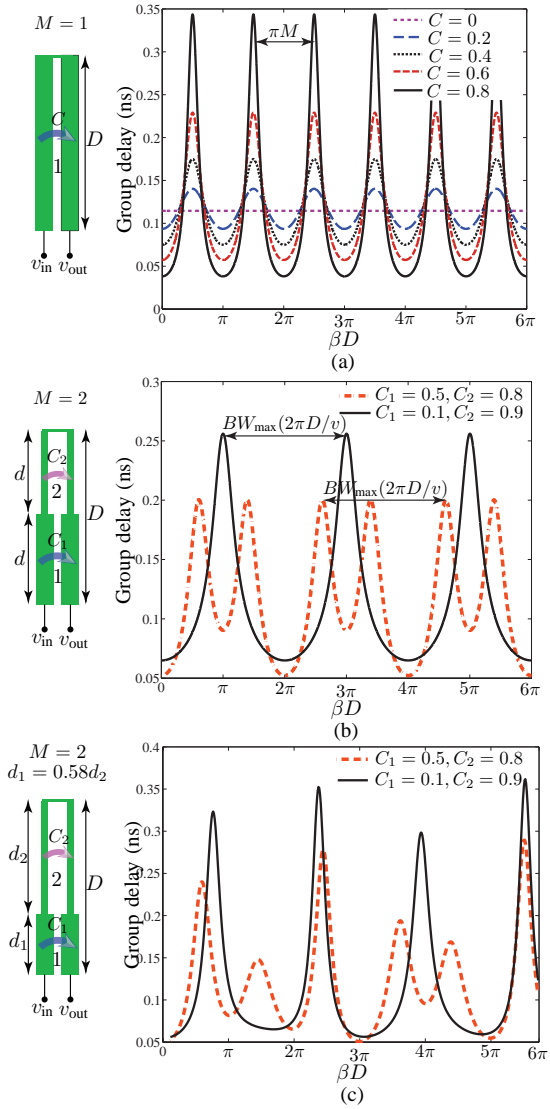}}}
\caption{Response of step-discontinuity nonuniform phasers~\cite{Schiffman_MTT_1955}. a)~Single C-section. b)~Two commensurate sections. c)~Two non-commensurate sections.}
\label{Fig:csections}
\end{figure}

Figure~\ref{Fig:csections}(a) shows that the group delay response of a single C-section is periodic, with peaks located at $\beta D=\pi(n+1/2)$, for $n=0,1,...,\infty$ and having a group delay swing depending on the coupling, $C$, and length, $D$, of the structure. Figure~\ref{Fig:csections}(b) shows that in a commensurate cascaded $M$-section C-section, the periodicity is increased by a factor~$M$ ($M$ propagation-coupled resonators) with up to $M$ peaks depending on couplings, due to \emph{coherent} multiple reflection [factor $e^{-j2m\beta d}$ in Eq.~\eqref{eqa:s21}]. Defining $BW_\text{max}$ as the frequency bandwidth supporting a non-periodic specified group delay response (restricted by periodicity), one has from $2\beta d=2\pi$ where $d=D/M$ that $BW_\text{max}=Mv/4D$. This reveals that the bandwidth of the phaser is increased by increasing $M$. Finally, Fig.~\ref{Fig:csections}(c) shows that periodicity is lost in the case of non-commensurate sections, due to \emph{coherent} multiple reflection [factor $e^{-j2m\beta d_m}$ in Eq.~\eqref{eqa:s21}].
\section{Continuously Modulated Nonuniform Phasers}\label{sec:cont_mod_nu}
To synthesize the nonuniform coupled-line function $C(z)$ ($0 <C_\text{min}<C(z)<C_\text{max}<1$) for the specified group delay response, one may use the Fourier series expansion
\begin{equation}
C(z)=a_0+\sum_{q=1}^Q \left[a_q \cos(2\pi qz/d)+b_q \sin(2\pi qz/d)\right],
\label{eqa:nu_coupling}
\end{equation}
and search for the appropriate unknown expansion coefficients $a_q$ and $b_q$. The corresponding nonuniform even and odd characteristic impedances are $Z_{0{e/o}}(z)=Z_0(\sqrt{1\pm C(z))/(1\mp C(z))}$~\cite{Pozar_4th_2012}. We shall satisfy the local matching condition, $\sqrt{Z_{0\text{e}}(z)Z_{0\text{o}}(z)}=Z_0,\quad\forall z$, where $Z_0$ is the ports characteristic impedance. The even and odd impedances at the input of the $m^\text{th}$ subsection, $Z^{\text{in}}_{\text{p},m}$, are related to those of the $(m+1)^\text{th}$ subsection, $Z^{\text{in}}_{p,m+1}$, by
\begin{equation}
Z_{p,m}^{\text{in}}=Z_{p,m}   \frac{Z_{p,m+1}^{\text{in}}+jZ_{p,m} \tan(\beta d_m)}{Z_{p,m}+jZ_{p,m+1}^{\text{in}} \tan(\beta d_m)}.
\end{equation}
Iteratively computing $Z^{\text{in}}_{p,m}$ from $m=M$ to $m=1$ provides the even and odd reflection coefficients at the input of the overall even and odd structures via $\Gamma_{p}^\text{in}=(Z^{\text{in}}_{{p},1}-Z_0)/(Z^{\text{in}}_{{p},1}+Z_0)$. The corresponding group delay of the phaser follows using~\eqref{eqa:gd}, and is injected into the fitness function $F=1/(\omega_\text{h}-\omega_{l})\int_{\omega_\text{l}}^{\omega_\text{h}} \left|\tau(\omega)-\tau_{\text{s}(\omega)}\right| d \omega$, for alignment of $\tau(\omega)$ with the specified function $\tau_{\text{s}}(\omega)$.

\begin{figure}
\centering
{\psfrag{a}[c][c][0.9]{(a)}
\psfrag{b}[c][c][0.9]{(b)}
\psfrag{A}[c][c][0.9]{Frequency (GHz)}
\psfrag{B}[c][c][0.9]{Group delay (ns)}
\psfrag{C}[l][c][0.7]{Specification, 1 to 5 GHz}
\psfrag{c}[l][c][0.7]{Specification, 1 to 20 GHz}
\psfrag{D}[l][c][0.7]{Continuously nonuniform}
\psfrag{E}[l][c][0.7]{Commensurate, $d_m=D/10$}
\psfrag{F}[l][c][0.7]{Non-commensurate}
\psfrag{q}[c][c][0.7]{\shortstack{BW restricting\\peaks}}
\psfrag{h}[c][c][0.75]{$d_1$}
\psfrag{i}[c][c][0.75]{$d_2$}
\psfrag{j}[c][c][0.75]{$d_3$}
\psfrag{k}[c][c][0.75]{$d_{10}$}
\psfrag{d}[c][c][0.8]{$1$}
\psfrag{e}[c][c][0.8]{$2$}
\psfrag{f}[c][c][0.8]{$3$}
\psfrag{g}[c][c][0.8]{$10$}
\psfrag{p}[c][c][0.8]{$D$}
\psfrag{x}[c][c][0.95]{$v_\text{in}$}
\psfrag{y}[c][c][0.95]{$v_\text{out}$}
\psfrag{Q}[c][c][0.7]{$\beta D=\pi$}
\centering{ \includegraphics[width=0.9\columnwidth] {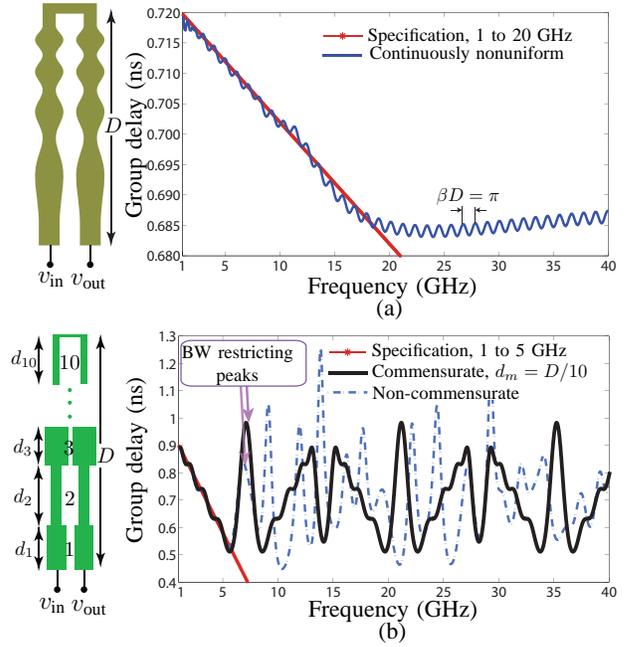}}}
\caption{Realization of a broad-band phasing (negative linear chirp) specification ($D=37~\text{mm}=0.353\lambda_{g,1 \text{GHz}}$). (a)~Continuously nonuniform phaser, covering the 1 to 20~GHz bandwidth. (b)~Step-discontinuity nonuniform commensurate and non-commensurate phasers with 10~sections, restricted to the 1 to 5~GHz bandwidth. }
\label{Fig:BW_enh}
\end{figure}

\begin{figure}[t!]
\psfrag{A}[c][c][0.8]{Frequency (GHz)}
\psfrag{B}[c][c][0.8]{Group Delay (ns)}
\psfrag{C}[l][c][0.7]{Specification}
\psfrag{D}[l][c][0.7]{Theory}
\psfrag{E}[c][c][0.7]{$D$}
\psfrag{F}[l][c][0.7]{$D_1$}
\psfrag{G}[l][c][0.7]{$D_2$}
\centering{ \includegraphics[width=1\columnwidth]{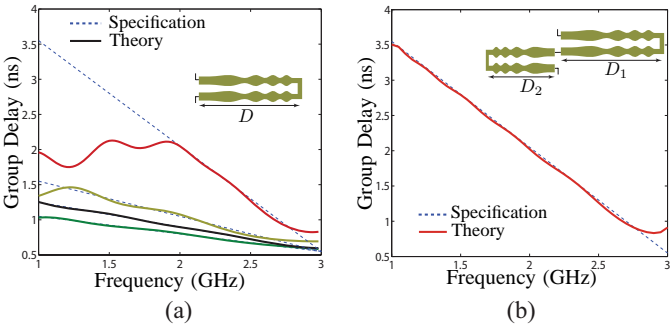}}
\caption{Nonuniform C-section cascading solution for suppressed oscillations and larger group delay swing. a)~Failure of a single continuously nonuniform C-section to reach $\Delta\tau>1$~ns. b)~Resolution of issue mentioned in a) by cascading two C-sections ($D_1=82$~mm and $D_2=52$~mm).}
\label{Fig:limitation_test}
\end{figure}

Figure~\ref{Fig:BW_enh} compares the performance of the continuously modulated nonuniform phaser with those of step-discontinuity nonuniform phasers. The goal is to a achieve negative linearly chirped response of at least $30$~ps swing over the largest possible bandwidth between 1 and 20~GHz\footnote{Note that the area under the $\tau(\omega)$ curve is constant for a given length~$D$~\cite{Gupta_TMTT_03_2015}.}. Due to its zero subsection length ($d/\lambda\rightarrow 0$), the continuously modulated phaser exhibits, according to Sec.~\ref{sec:cascaded}, an infinite periodicity, and reaches therefore the complete specified bandwidth. In contrast, the bandwidth of the step-discontinuous phasers is restricted by spurious peaks due to excessive subsection length. The oscillations in the group delay curves, more visible in the continuously nonuniform case due to smaller large-scale variations, correspond to the resonances of the overall C-section structures ($\beta D=\pi$, i.e. $\Delta f=1.11$~GHz). These oscillations may be suppressed by using cascaded non-uniform C-sections, which also allows to increase the group delay swing, as shown in Fig.~\ref{Fig:limitation_test}.

\begin{figure}
\begin{center}
\subfigure[]{\label{Fig:coupl_gd_cheb_a}
\psfrag{A}[c][c][0.95]{$z/d$}
\psfrag{B}[c][c][0.95]{Coupling $C$}
\psfrag{C}[l][c][0.85]{Cheb. 1$^\text{st}$ order}
\psfrag{D}[l][c][0.85]{Cheb. 2$^\text{nd}$ order}
\psfrag{E}[l][c][0.85]{Cheb. 3$^\text{rd}$ order}
\psfrag{F}[l][c][0.85]{Cheb. 4$^\text{th}$ order}
\includegraphics[width=0.76\columnwidth]{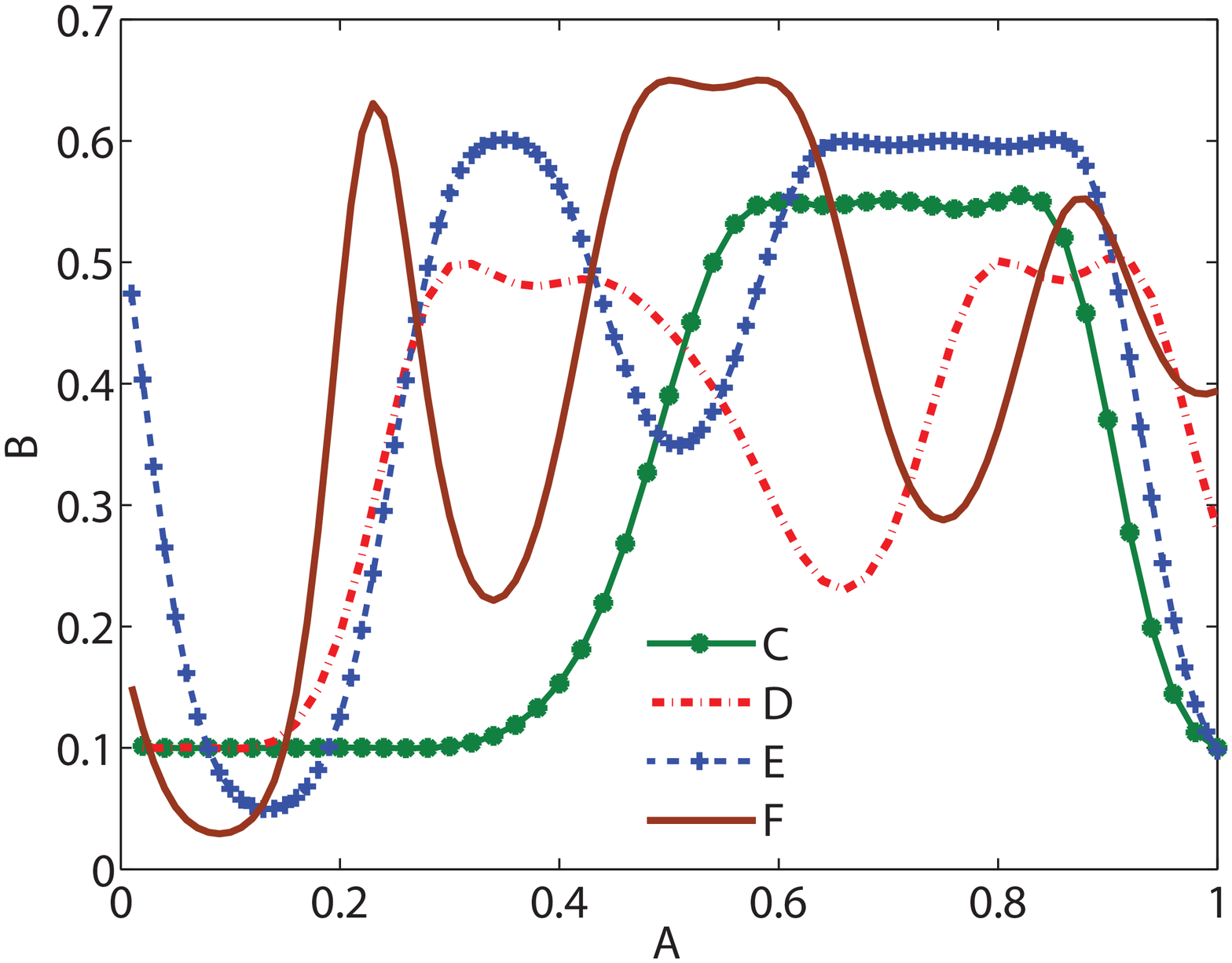}\label{Fig:coupling_compare_theory_1st_2nd_3rd}}
\subfigure[]{\label{Fig:coupl_gd_cheb_b}
\psfrag{A}[c][c][0.95]{Frequency (GHz)}
\psfrag{B}[c][c][0.95]{Group Delay $\tau$ (ns)}
\psfrag{C}[l][c][0.85]{Specifications}
\psfrag{D}[l][c][0.85]{Cheb. 1$^\text{st}$}
\psfrag{E}[l][c][0.85]{Cheb. 2$^\text{nd}$}
\psfrag{F}[l][c][0.85]{Cheb. 3$^\text{rd}$}
\psfrag{G}[l][c][0.85]{Cheb. 4$^\text{th}$}
\psfrag{H}[l][c][0.85]{Uniform}
\includegraphics[width=0.76\columnwidth]{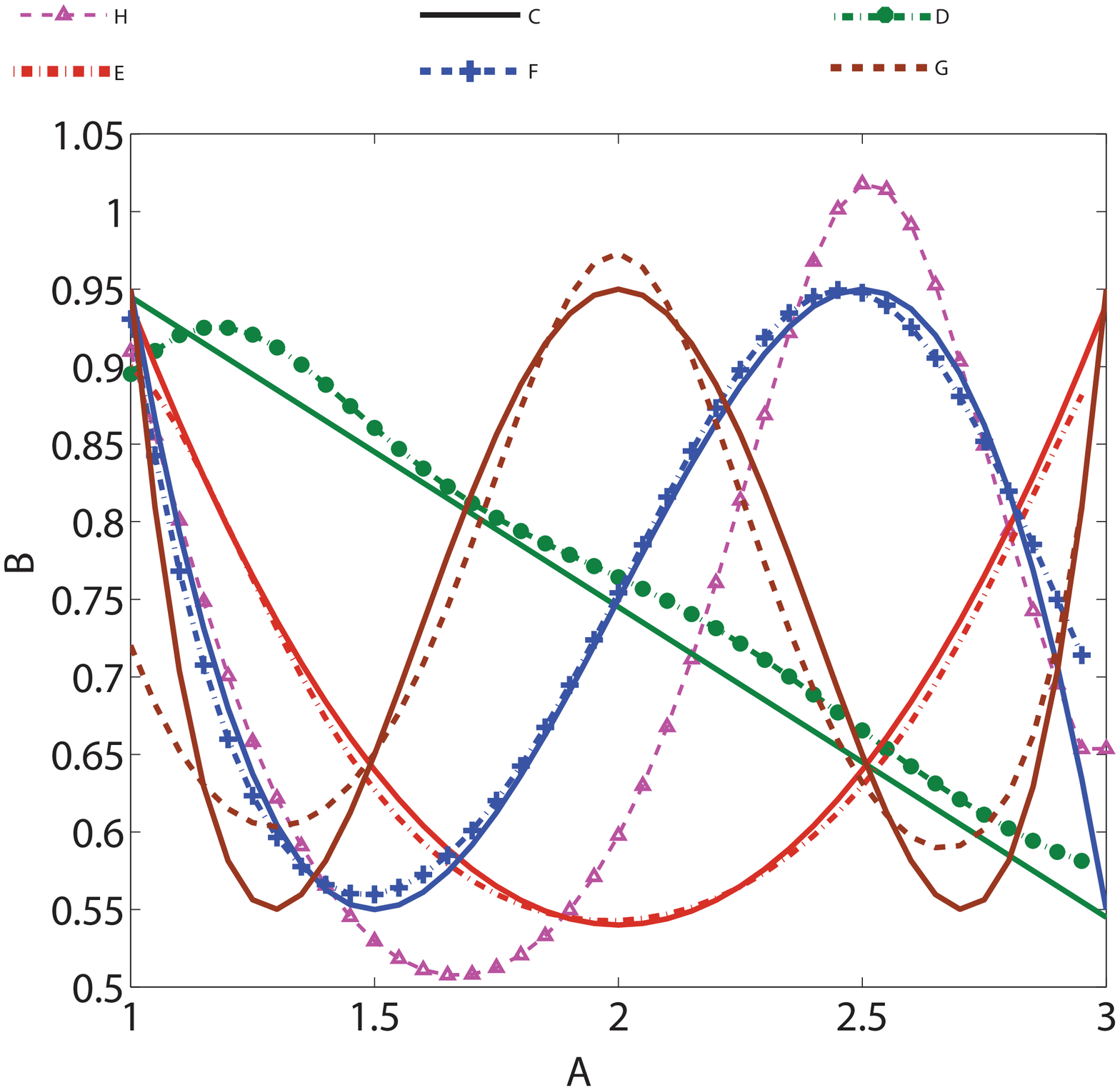}\label{Fig:gd_compare_theory_1st_2nd_3rd_4th}}
\caption{Realization of Chebyshev first four orders group delay responses, using the synthesis technique presented at the beginning of Sec.~\ref{sec:cont_mod_nu}. a)~Nonuniform coupling function $C(z)$ [Eq.~\eqref{eqa:nu_coupling}]. b)~Group delay response [Eq.~\eqref{eqa:gd}].}
\label{Fig:coupl_gd_cheb}
\end{center}
\end{figure}

\begin{figure}
\begin{center}
\subfigure[]{\label{Fig:1st_meas_sim_theory_gd}
\psfrag{A}[c][c][0.8]{Frequency (GHz)}
\psfrag{B}[c][c][0.8]{Group Delay (ns)}
\psfrag{C}[l][c][0.6]{Specification}
\psfrag{D}[l][c][0.6]{Theory}
\psfrag{E}[l][c][0.6]{Full-Wave}
\psfrag{F}[l][c][0.6]{Measured}
\includegraphics[width=0.47\columnwidth]{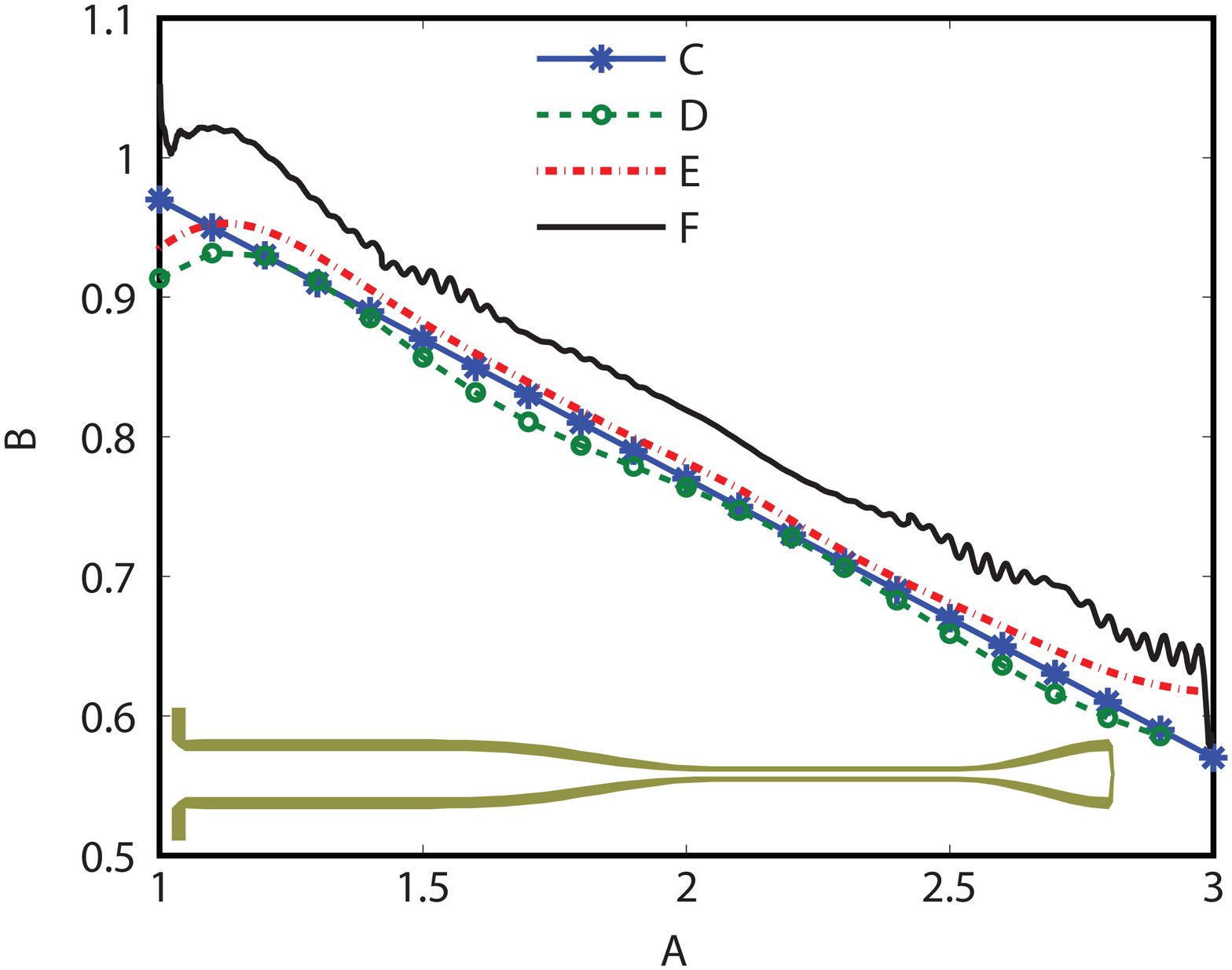}}
\subfigure[]{\label{Fig:1st_meas_sim_theory_sp}
\psfrag{A}[c][c][0.8]{Frequency (GHz)}
\psfrag{B}[c][c][0.8]{S-Parameters (dB)}
\psfrag{C}[l][c][0.6]{$S_{11}$-Full-Wave}
\psfrag{D}[l][c][0.6]{$S_{21}$-Full-Wave}
\psfrag{E}[l][c][0.6]{$S_{11}$-Measured}
\psfrag{F}[l][c][0.6]{$S_{21}$-Measured}
\includegraphics[width=0.47\columnwidth]{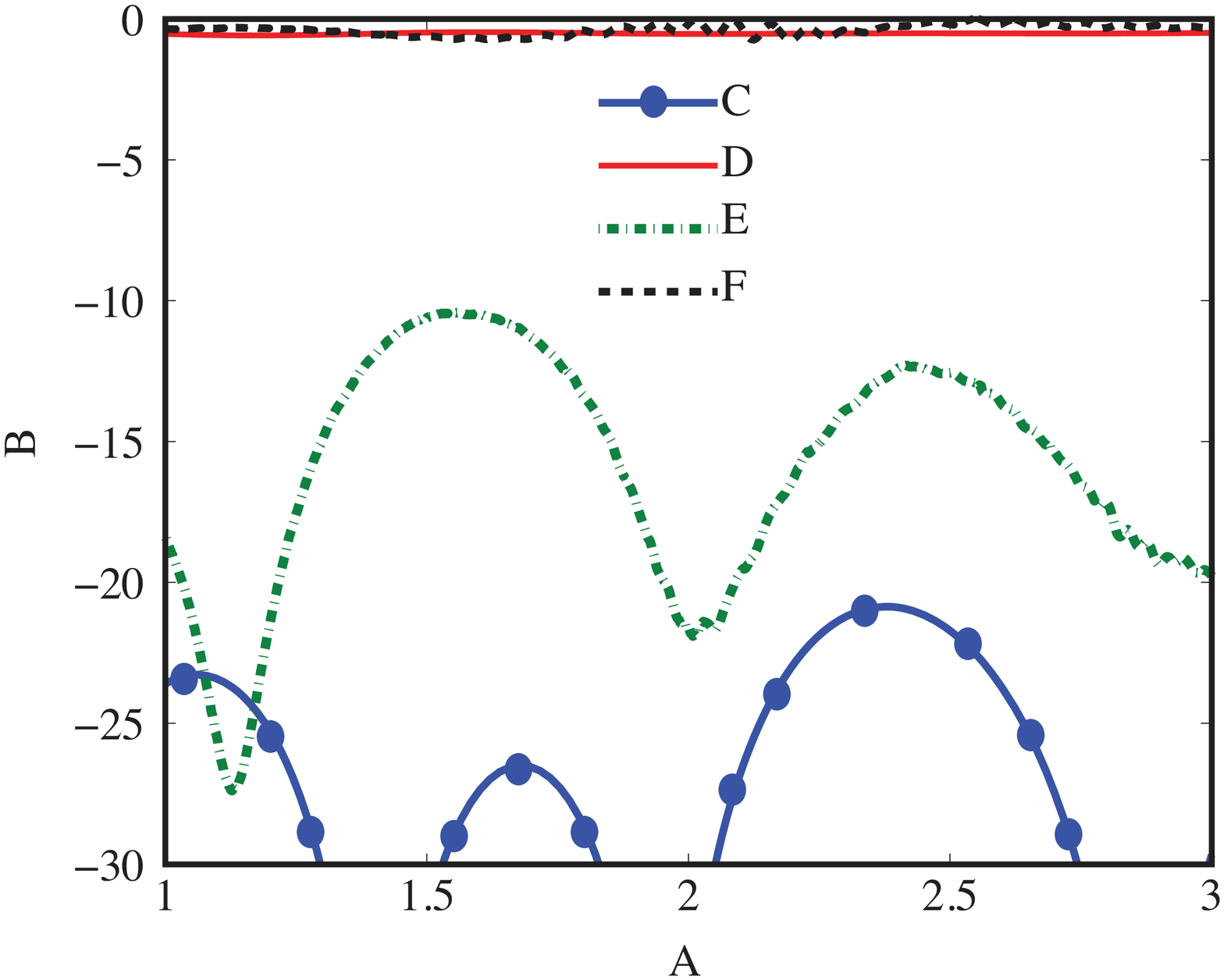}}
\subfigure[]{\label{Fig:2nd_meas_sim_theory_gd}
\psfrag{A}[c][c][0.8]{Frequency (GHz)}
\psfrag{B}[c][c][0.8]{Group Delay (ns)}
\psfrag{C}[l][c][0.6]{Specification}
\psfrag{D}[l][c][0.6]{Theory}
\psfrag{E}[l][c][0.6]{Full-Wave}
\psfrag{F}[l][c][0.6]{Measured}
\includegraphics[width=0.47\columnwidth]{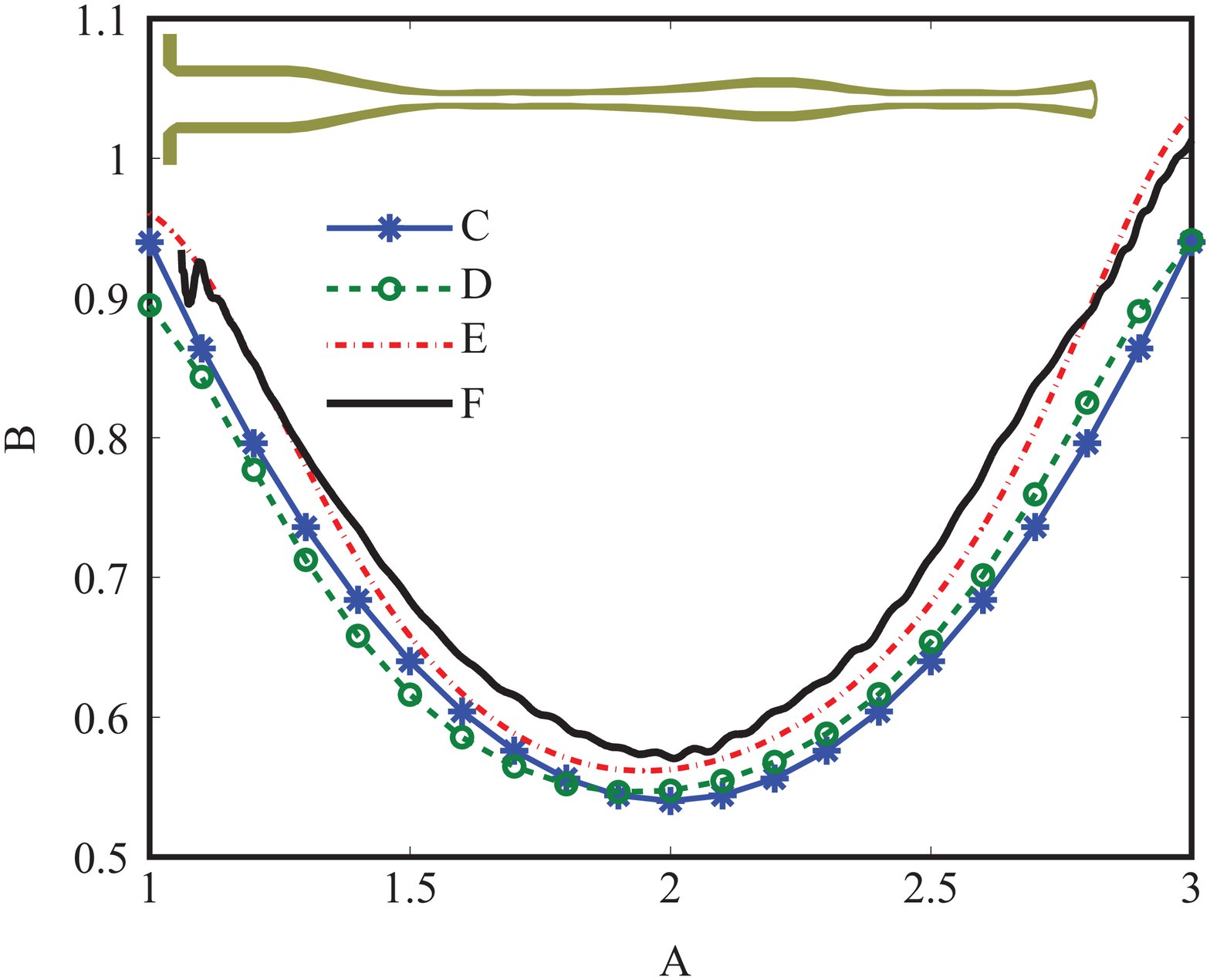}}
\subfigure[]{\label{Fig:2nd_meas_sim_theory_sp}
\psfrag{A}[c][c][0.8]{Frequency (GHz)}
\psfrag{B}[c][c][0.8]{S-Parameters (dB)}
\psfrag{C}[l][c][0.6]{$S_{11}$-Full-Wave}
\psfrag{D}[l][c][0.6]{$S_{21}$-Full-Wave}
\psfrag{E}[l][c][0.6]{$S_{11}$-Measured}
\psfrag{F}[l][c][0.6]{$S_{21}$-Measured}
\includegraphics[width=0.47\columnwidth]{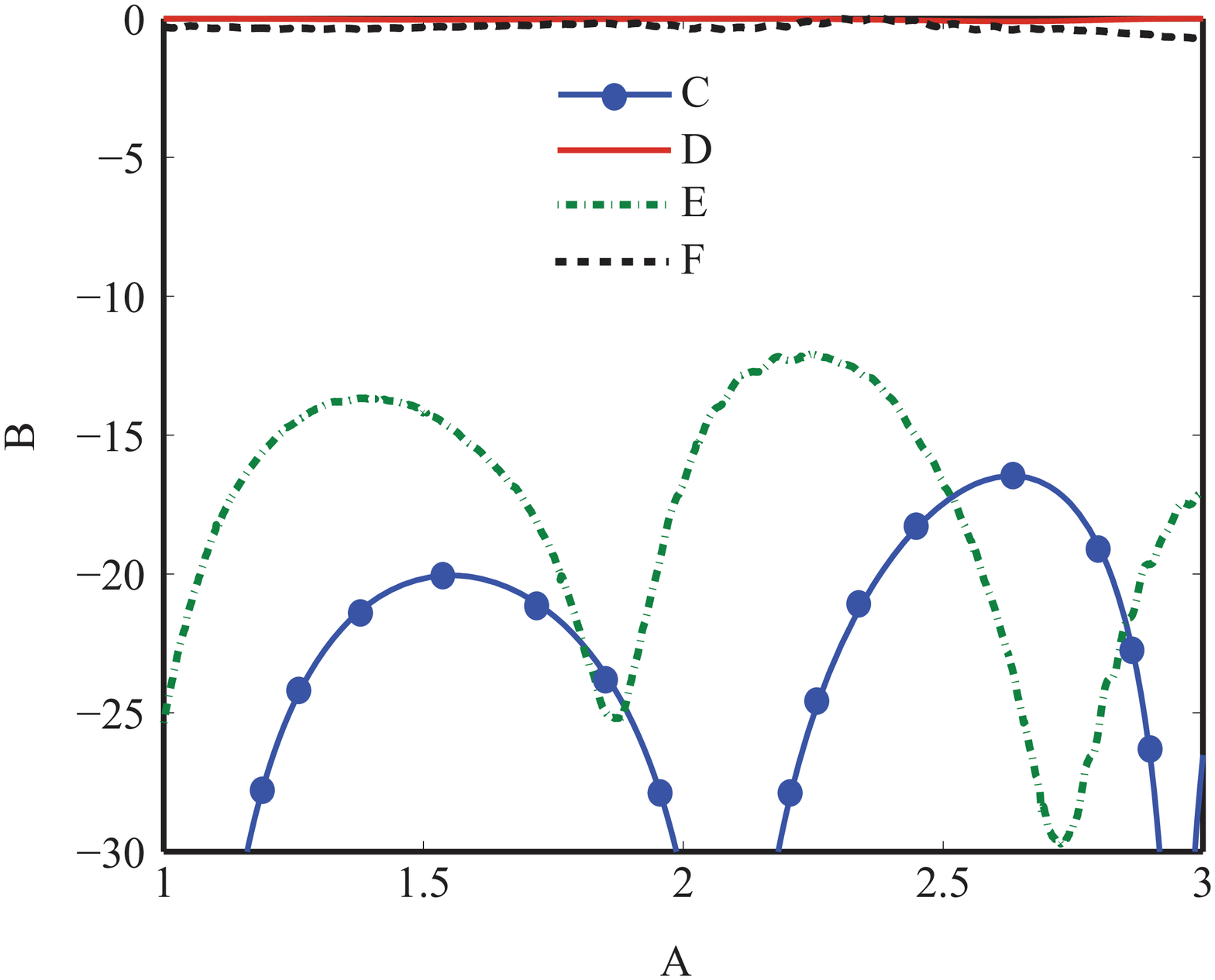}}
\caption{Experimental results compared with full-wave results (CST Microwave Studio) for Chebyshev phaser in stripline technology (layouts shown as insets). a)~$1^\text{st}$ order group delay. b)~$1^\text{st}$ order S-parameters. c)~$2^\text{nd}$ order group delay. d)~$2^\text{nd}$ order S-parameters.}
\label{Fig:gd_sp_1st_comparison}
\end{center}
\end{figure}

Let us finally demonstrate the dispersion diversity of the phaser by specifying 1$^\text{st}$ order to 4$^\text{th}$ order Chebyshev group delay responses. This benefits comes from the virtually unlimited degrees of freedom of the continuously nonuniform structure, in contrast to its super-wavelength step-discontinuity counterparts. Illustrative results are shown in Fig.~\ref{Fig:coupl_gd_cheb} for Chebyshev group delay specifications, while experimental validations for the $1^\text{st}$ and $2^\text{nd}$ orders are presented in Fig.~\ref{Fig:gd_sp_1st_comparison}.
\section{Conclusions}

We have shown that a continuously nonuniform coupled-line C-section phaser provides enhanced bandwidth and profile diversity compared to step-discontinuity coupled-line C-section phasers. Such a phaser, that may be further cascaded for oscillation suppression and delay swing enhancement, is a promising device for in real-time analog signal processing (R-ASP). It may be for instance applied to Dispersion Code Multiple Access (DCMA), a recently proposed novel multiplexing wireless technology~\cite{Nikfal_Patent_2013}, where the number of channels is equal to the number of available phaser responses and cross-channel interference is minimized using Chebyshev dispersion profiles.
\bibliographystyle{IEEEtran}
\bibliography{Travati_NonuniformCsection_MWCL_Reference}

\begin{thebibliography}{1}
\providecommand{\url}[1]{#1}
\csname url@samestyle\endcsname
\providecommand{\newblock}{\relax}
\providecommand{\bibinfo}[2]{#2}
\providecommand{\BIBentrySTDinterwordspacing}{\spaceskip=0pt\relax}
\providecommand{\BIBentryALTinterwordstretchfactor}{4}
\providecommand{\BIBentryALTinterwordspacing}{\spaceskip=\fontdimen2\font plus
\BIBentryALTinterwordstretchfactor\fontdimen3\font minus
  \fontdimen4\font\relax}
\providecommand{\BIBforeignlanguage}[2]{{%
\expandafter\ifx\csname l@#1\endcsname\relax
\typeout{** WARNING: IEEEtran.bst: No hyphenation pattern has been}%
\typeout{** loaded for the language `#1'. Using the pattern for}%
\typeout{** the default language instead.}%
\else
\language=\csname l@#1\endcsname
\fi
#2}}
\providecommand{\BIBdecl}{\relax}
\BIBdecl

\bibitem{Caloz_MM_RASP_Caloz_09_2013}
C.~Caloz, S.~Gupta, Q.~Zhang, and B.~Nikfal, ``Analog signal processing,''
  \emph{{Microw. Mag.}}, vol.~14, no.~6, pp. 87--103, Sept. 2013, invited.

\bibitem{Gupta_TMTT_03_2015}
S.~Gupta, Q.~Zhang, L.~Zou, L.~J. Kiang, and C.~Caloz, ``Generalized
  coupled-line all-pass phasers,'' \emph{{IEEE Trans. Microw. Theory Techn.}},
  vol.~63, no.~3, pp. 1007--1018, Mar. 2015.

\bibitem{Laso_MTT_60_2012}
M.~Chudzik, I.~Arnedo, A.~Lujambio, I.~Arregui, I.~Gardeta, F.~Teberio, J.~A.
  {n}a, D.~Benito, M.~A.~G. Laso, and T.~Lopetegi, ``Design of
  transmission-type nth-order differentiators in planar microwave technology,''
  vol.~60, no.~11, pp. 3384--3394, Nov. 2012.

\bibitem{Nikfal_MWCL_11_2012}
B.~Nikfal, D.~Badiere, M.~Repeta, B.~Deforge, S.~Gupta, and C.~Caloz,
  ``Distortion-less real-time spectrum sniffing based on a stepped group-delay
  phaser,'' \emph{{IEEE Microw. Wireless Compon. Lett.}}, vol.~22, no.~11, pp.
  601--603, Oct. 2012.

\bibitem{Nikfal_Patent_2013}
B.~Nikfal, M.~Salem, and C.~Caloz, ``A method and apparatus for encoding data
  using instantaneous frequency dispersion,'' Patent US 62/002,978, Nov., 2013.

\bibitem{Taravati_EuMC_2014}
S.~Taravati, Q.~Zhang, and C.~Caloz, ``Non-uniform {C}-section phasers for
  enhanced design flexibility in radio analog signal processing,'' in
  \emph{{IEEE European Microw. Conf. (EuMC)}}, Rome, Italy, Oct. 2014.

\bibitem{Schiffman_MTT_1955}
B.~M. Schiffman, ``A new class of broad-band microwave 90-degree phase
  shifters,'' vol.~3, no.~2, pp. 232 -- 237, Oct. 1955.

\bibitem{Cristal_TMTT_1966}
E.~G. Cristal, ``Analysis and exact synthesis of cascaded commensurate
  transmission~line c-section all~pass networks,'' \emph{{IEEE Trans. Microw.
  Theory Techn.}}, vol.~14, no.~6, pp. 285--291, Jun. 1966.

\bibitem{Pozar_4th_2012}
D.~M. Pozar, \emph{Microwave Engineering}, 4th~ed.\hskip 1em plus 0.5em minus
  0.4em\relax John Wiley \& Sons, Inc., 2012.

\end{thebibliography}
\end{document}